\documentclass[useAMS,usenatbib]{pasa}

\title[Light on Dark Matter]{Light on Dark Matter}
\author[Tully]{R. Brent Tully,$^{1,2}$}
\affil{
$^1$Institute for Astronomy,University of Hawaii, 2680 Woodlawn Drive, 
Honolulu, HI 96822, USA\\
$^2$tully@ifa.hawaii.edu}
\begin{document}

\maketitle

\label{firstpage}

\begin{abstract}
Galaxies are lighthouses that sit atop peaks in the density field.
There is good observational evidence that these lighthouses do not provide
a uniform description of the distribution of dark matter.

\end{abstract}

\begin{keywords}
dark matter-galaxies: clusters-galaxies: dwarf-galaxies:luminosity function,
mass function
\end{keywords}

\section{Let there be Light}

In the beginning there was dark matter and gas but there were no stars.
Today some of the gas that was once dispersed has transformed into stars,
and we have light.  In places the stars are young, hot, and bright while 
elsewhere the stars are old, cool and faint.  In places the stars have been 
scattered by violent collisions.  In places the gas never coalesced and 
stars never formed.  We have light, but light in selective places.

It has been pointed out that the abrupt cutoff of the luminosity function
at the bright end and the flat slope at the faint end compared with the 
Press-Schechter mass function anticipated by the hierarchical clustering
scenario suggests that there is a relative deficiency of light at high and
low mass extremes compared with the situation at intermediate masses
(Yang et al. 2003, van den Bosch et al. 2003).  Semi-analytic models that
follow the transformation of gas into stars lead to similar expectations
(Blanton et al. 1999, Somerville et al. 2001, Ostriker et al. 2003).  These
same results are found directly in observations. A full discussion of the
observational situation is provided by Tully (2004).  A condensed version
is presented here.

\section{The High End of the Mass Spectrum}

Two lines of evidence strongly indicate that the ratio of blue light
to mass {\it decreases by about a factor 6 in progressing from halos of mass
$10^{12} M_{\odot}$ to halos of mass $10^{15} M_{\odot}$.}

\subsection{Virial Analysis of Groups}

\begin{figure}
\includegraphics{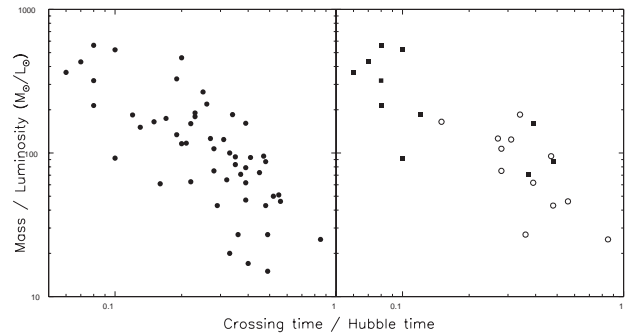}
\vspace*{45mm}
\caption{Correlation between mass/light ratio and group crossing time.
{\it Left panel:} all groups with 5 or more members in the volume limited
T87 sample.  {\it Right panel:} Subset of groups with at least 6 members
with $M_B<-17$ distinguished according to whether a majority are E-S0-Sa
(filled squares) or Sab-Sc-Irr (open circles).
\label{fig1}}
\end{figure}

Collapsed halos are the sites of groups and clusters of galaxies (where in
some circumstances the `group' may be a single object).  The relative 
distribution and motions of the galaxies in a group or cluster provide an 
indication of the mass of the ensemble.  Many groups are 
relatively young (crossing times are a substantial fraction of the age of 
the universe) so are not expected to be relaxed.  As a consequence, mass
estimations based on the virial theorem have uncertainties of
as much as a factor 2.  Nonetheless, bear with this analysis 
based on the virial assumption because the variances in mass to light with
environment that will be demonstrated are much larger than a factor 2.

A volume limited sample of groups in the Local Supercluster, extending to
$25 h_{75}^{-1}$ Mpc ($h_{75} = {\rm H}_0 / 75$), was assembled by Tully 
(1987: T87), with the individual galaxies in the T87 groups identified 
in Table II of Tully (1988).  The base galaxy catalog was considered 
complete to $M_B^{\star}$ within this volume.
The group catalog consists of 179 groups of
two or more galaxies, 50 with five or more.  

The left panel of Figure 1 reveals a striking correlation between the 
virial mass $M_V$ to blue light $M_B$ ratio and the dynamical timescale 
of groups
as measured by crossing time $t_X$.  There is an increase of an order of 
magnitude in $M_V/L_B$ as $t_X$ decreases from $\sim 0.5 {\rm H}_0$ to
$<0.1 {\rm H}_0$.  The right panel of the figure shows a subset of the 
sample, groups with at least 6 non-dwarf members, distinguished by 
whether the majority of the members are early or late type.  There is a
clear separation: the spiral rich groups have long dynamical times and
low $M_V/L_B$ values compared with the groups dominated by E/S0 galaxy
types.

Mass and luminosity are separated in Fig.~2.  It is seen that the
E/S0/Sa dominated groups tend to be the most massive.  Two concurrent
trends are found: more massive groups tend to have fractionally
less blue light, and at a given mass E/S0/Sa groups have less blue light
than their counterparts dominated by later morphological types.  

The properties of groups from the volume limited survey region with only 
2--4 known members are included in Fig.~2 in order to explore conditions in
a lower mass range.  In these cases the errors in the mass estimates are very
large, which accounts for the large scatter.  The 5 data points with error bars
in this figure are discussed later.

\begin{figure}
\includegraphics{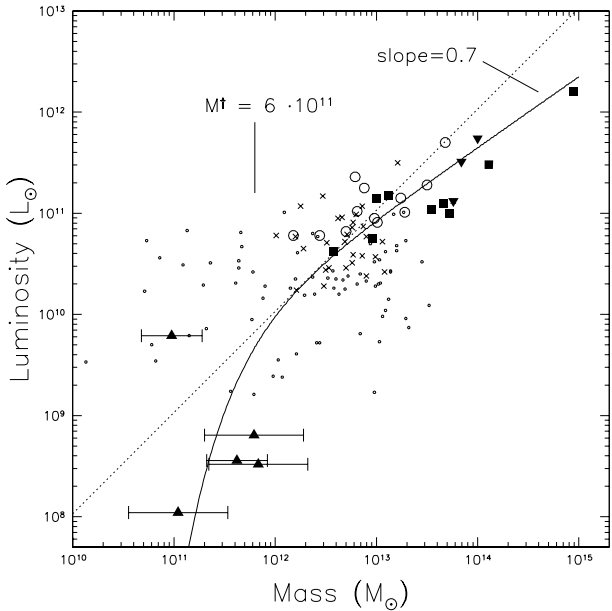}
\vspace*{85mm}
\caption{
Mass vs. light for groups over the full range of the observed mass spectrum.  
The large open circles and filled squares represent the same data seen in the 
right panel of Fig.~1.  The crosses represent the remainder of the groups with
5 or more members found in the left panel of Fig.~1.  The small open circles 
represent groups with only 2--4 known members.  The 3 filled inverted triangles
represent E/S0/Sa groups more distant than the volume limited sample.  The 5
filled triangles with error bars represent groups at the low end of the mass
spectrum that are discussed separately.
The $M_V/L_B=94$ dotted line is the mean value for the sample reported by T87. 
The solid line is a fit constrained by the data with a high mass slope of 
$\gamma=0.7$
and an exponential fall off at masses less than
$M^{\dag} = 6 \times 10^{11} M_{\odot}$.
\label{fig2}}
\end{figure}

\subsection{Numerical Action Methods}

\begin{figure}
\includegraphics{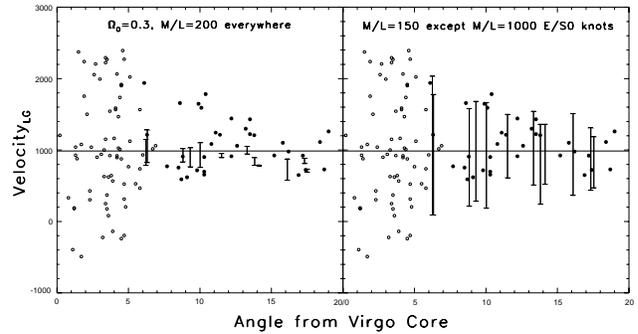}
\vspace*{48mm}
\caption {Virgo infall constraints from two Numerical Action models.
Data points indicate the velocities and separations from the center of the
Virgo Cluster of individual galaxies.  Galaxies within the $6^{\circ}$
caustic of the cluster are indicated by open circles.  Galaxies outside the
caustic but associated with the infall region are indicated by filled circles.
The vertical brackets are located at angles from the center of the cluster
that intersect infalling groups, so at lines-of-sight that have received
attention in the Numerical Action models.  The amplitudes of the brackets
illustrate the range of infall velocities anticipated by the two models
under consideration. {\it Left panel:} $M/L_B=200$ assigned to all units.
{\it Right panel:} $M/L_B=1000$ assigned to the Virgo Cluster and other 
E/S0 knots and $M/L_B=150$ otherwise.
\label{fig3}}
\end{figure}

Compatible results have emerged from the reconstruction of galaxy orbits with
Numerical Action Methods (Peebles 1989, Shaya et al. 1995, Tully and Shaya 
1998).  These reconstructions extend dynamical studies to larger scales.
One is considering the dynamical influence of groups on each other and 
potentially taking into account mass that extends beyond the groups.
For this discussion, no attempt is made to model the complex orbits 
{\it within} groups.  Groups are considered as ensemble halos and it is 
the orbits of the centers of mass of these ensembles that are reconstructed.

In fact it is not the specific orbits that are reconstructed that we take 
seriously but rather it is the mass requirements that are found to give good
fits.  The reconstruction is constrained by the need to match the
angular positions and radial velocities of objects in a catalog.  If the
cosmology is set with a specified age then the main variable is the mass
assignments.  With a given set of mass assignments, orbits are found which
leaves each entity today with 6 specified values in phase space, 3 of which
rigorously match the observed boundary conditions.  One of the other 3
phase space values -- the distance -- is subject to test.  Our model constraint
is a $\chi^2$ comparison of model and observed distances.  The orbits that
are found are not unique given the limitations of the catalogs and the
poor constraints on the 3 free phase space components, but {\it only models
within a modest range of mass assignments give good $\chi^2$ fits to the
distance constraints.}  Basically, increasing mass forces objects with a
given velocity to be farther removed from the main mass centers.

In the earliest trials the minimalist assumption was made that $M/L_B$ values
are constant in all environments.  The subsequent best fits were found for
$M/L_B \sim 200 M_{\odot}/L_{\odot}$.  However this modest $M/L_B$ assignment
fails utterly to account for the large infall velocities toward the Virgo
Cluster that are observed.  Rather, a value 
$M/L_B \sim 1000 M_{\odot}/L_{\odot}$ is required for the cluster.
Figure~3 attempts to summarize the observational constraints.  The amplitude
of the envelope of observed infall velocities is closely governed by the
mass of the cluster.  A very high mass of $1.3 \times 10^{15} M_{\odot}$
is required to explain the velocity breadth of the infall region. Upon
assigning the corresponding large mass to light ratio to the cluster 
(and other E/S0 dominated groups)
it was found that the mass to light requirement for the remainder of the 
objects was reduced to $M/L_B \sim 150 M_{\odot}/L_{\odot}$. 

The mass requirement found by the Numerical Action Method for the Virgo
Cluster is 60\% higher than found with the virial analysis, but applies
to the cluster on a larger scale.  The overall average $M/L_B$ requirement
of the action modelling is also higher than found via the virial theorem for 
most groups but
again refers to larger scales.  Both the virial and action studies suggest
that $M/L_B$ values increase by a factor 5--7 in proceeding from the
environments of spiral galaxies to the Virgo Cluster environment. 

\section{The Low End of the Mass Spectrum}

The faint end of the luminosity function of galaxies in low density
environments has a substantially flatter slope than the theoretical
mass function of hierarchical clustering (Klypin et al. 1999, Moore et al.
1999).  During a study of possible variations of the luminosity function
with environment (Tully et al. 2002) the idea arose that groups with low
mass with very little light might be common.  It was recalled that during 
the construction of the T87 group catalog a number of entities that were
called `associations' had group-like qualities except in one respect.
These entities had similar, though somewhat smaller, dimensions to loose 
groups and similar, actually smaller, velocity dispersions.  If considered
to be bound, then the inferred masses are low -- in the range 
$10^{11-12} M_{\odot}$.  If one ignores the properties of the galaxies and
regards them only as test particles probing a potential then one would 
conclude that these T87 associations are just the low mass end of a continuum
of groups conditions.

The outstanding difference with familiar groups with higher mass is that 
these low mass entities have {\it very little light.}  The candidate members
of these groups are all dwarf galaxies.  Though the inferred masses are low, 
the luminosities are {\it extremely low} and $M/L_B$ values are high -- so 
high that at the time of T87 there was a reluctance to conclude that the 
entities are bound (though there was a suspicion that this was the case!)

Today, with greater boldness, it is argued that these groups of dwarfs
{\it are} bound.  It is to be noted that candidate groups of dwarfs are not
rare.  Within the restricted domain of 5 Mpc, Tully et al. (2002) report 4 
good examples.  This number is comparable to the number of established groups
within the same distance, and the census of dwarfs is surely incomplete at
low Galactic latitudes.  As discussed by Tully (2004), there is now a 
dramatic improvement in knowledge regarding the distances of nearby galaxies.
Hubble Space Telescope observations resolves the brighter stars and permits a
distance determination from the magnitude of the tip of the red giant branch
(Karachentsev et al. 2003).  Good distances now exist for almost all the
dwarf group candidates and it is found that the spatial correlations are
equally as good in distance as they were found to be on the plane of the
sky and in velocity.  One of the groups of dwarfs turned out to be considerably
closer than suspected based on velocities, with all the observed distances 
concordant at the nearer position.

At this point the evidence very strongly suggests that the groups of dwarfs
are bound.  If so, mass estimates suggest $M/L_B \sim 1000 M_{\odot}/L_{\odot}$
for these groups.
The structures are almost surely not virialized because
crossing times are a substantial fraction of the Hubble time.  Consequently,
the derivation of masses are subject to large errors.  The uncertainties
can easily be a factor 2 and conceivably as bad as a factor 3.  However,
the $M/L_B$ values that are found are an order of magnitude larger than 
the values attributed to loose groups of spirals.  There is little doubt that
if the groups of dwarfs are bound then they are reservoirs of mostly dark 
matter.

The mass and light values of the groups of dwarfs are shown in Fig.~2 as the
filled triangles with error bars.  In fact, five points are plotted in this
manner and one stands apart in terms of luminosity.  This higher luminosity
entity is the near side component of what is commonly referred to as the 
Sculptor Group,but which is probably two separate bound units.  The near
component includes NGC 55 and NGC 300 plus smaller objects.  The dynamical
properties (dimensions and velocity dispersion) suggest a very low mass
for this proposed group, like that of the other 4 groups of dwarfs.  However,
the two NGC galaxies are intermediate luminosity galaxies, not dwarfs, so
the group is not deficient in light.  It is to be concluded that not all
group halos in the mass regime $10^{11-12} M_{\odot}$ lack light.
However in the ensemble there seems to be a cutoff in light proceeding to
low mass halos.

\section{The Dark Side}

The solid line in Fig.~2 is a fit of the equation
\begin{equation}
L_B = C M^{\gamma} e^{-M^{\dag}/M}
\end{equation}
This equation relates luminosity and mass with 3 constraints: a logarithmic
slope at the
high mass end, $\gamma$, a low mass exponential cutoff set by $M^{\dag}$,
and a normalization, $C$.  The fit minimizes a $\chi^2$ expression with
uncertainties in mass.  Perhaps a more dramatic presentation of the same 
information is seen in Figure~4, a plot of $M/L_B$ vs. mass (appreciating
that the groups with 2-4 members represented by the small open circles have
very large uncertainties in mass). 
It is seen that
halos in the mass range $10^{12-13} M_{\odot}$ manifest the most light.
Groups with progressively more mass up to $10^{15} M_{\odot}$ have 
progressively smaller light fractions, with this trend particularly enhanced
in the groups dominated by early type galaxies.  At the other end of the mass
spectrum, below $10^{12} M_{\odot}$ there is an abrupt cut-off, with 
$M/L_B$ values becoming very large and perhaps going off to infinity. One
can suppose that there are low mass halos completely devoid of stars.

\begin{figure}
\includegraphics{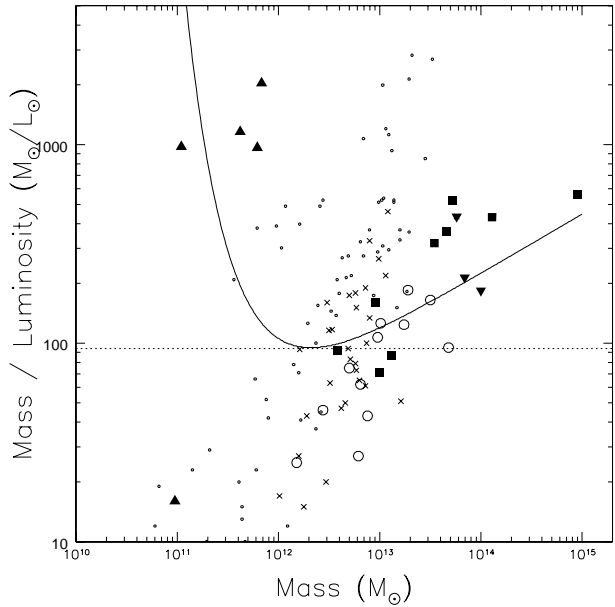}
\vspace*{83mm}
\caption{
$M/L_B$ vs. mass over the full range of density regimes.  The data
are the same as in Fig.~2.
The solid curve and dotted line are transpositions of the curve and line in
Fig.~2.
\label{fig4}}
\end{figure}

The census of galaxies within $25 h_{75}^{-1}$ Mpc of T87 that lead to the
group catalog can be used to construct a halo mass function.  This is seen 
in the top panel of Figure~5.  The mass function is poorly constrained at 
the high mass end by the small number statistics in such a restricted volume.
At the low mass end, the error bars only account for the $\sqrt{N}$ bin
counting uncertainties.  There are much larger horizontal errors associated 
with the uncertain masses of small groups.  In the lowest mass bins the
counts are dominated by individual galaxies outside of groups, in which cases
it was assumed that $M/L_B=100$.  

Admitting that there are considerable uncertainties in this mass function
because of the restricted sample and low mass cutoff, perhaps the greater 
interest is in the lower panel of Fig.~5.  Here one see the fraction of the 
total mass found in the separate mass bins.  The filled part of the histogram
is contributed by the E/S0/Sa groups, while the open part represents the
contribution by the groups dominated by spiral and irregular galaxies.
One entity, the Virgo Cluster, contains $\sim 40\%$ of the
mass in the survey volume (although it contains only 15\% of the light!). 
The groups of early types together contain 60\%
of the mass.  Considering groups of all type, 90\% of the mass is in groups
with $M > 3 \times 10^{12} M_{\odot}$.

\begin{figure}
\includegraphics{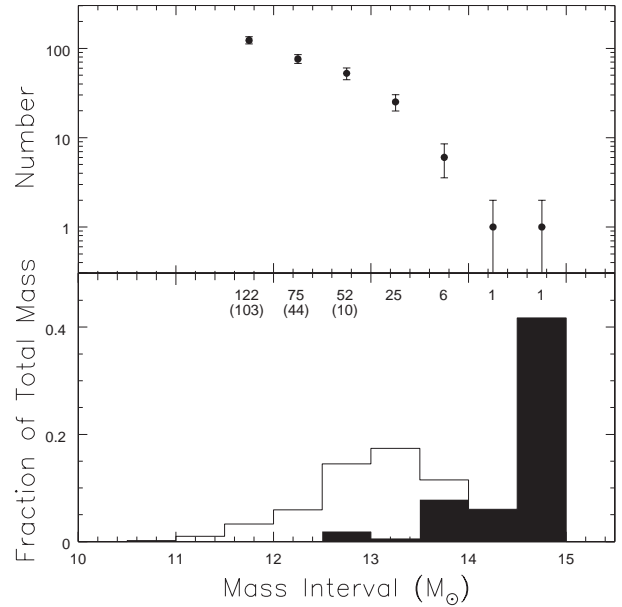}
\vspace*{85mm}
\caption{
{\it Top panel:} Observed mass function in the volume within
$25 h_{75}^{-1}$~Mpc
and $\mid b \mid > 30$.  The numbers of objects in each bin are given along the
top of the lower panel.  In brackets are the numbers of individual
galaxies; ie, galaxies outside of groups.  {\it Lower panel:} Fraction of the
total mass associated with the mass function in half-dec bins.  The mass
associated with the groups of predominantly early types is indicated by the
filled histograms.  The open histograms correspond to the mass found in the
groups with predominantly late types.
\label{fig5}}
\end{figure}

The groups of dwarf galaxies have only been discovered within a restricted 
5 Mpc volume.  Within this space, though they appear to be numerous, they only
contain $\sim 10\%$ of the mass associated with luminous galaxies.  Yet the
luminous groups within 5 Mpc lie on the tail of the histogram in Fig.~5 just
above $10^{12} M_{\odot}$.  It is concluded that these groups of dwarfs, though
they are dominated by dark matter, they make only a minor contribution to the
total mass budget of collapsed dark matter halos.  From a cosmological
perspective, it is the high mass to light ratios in massive clusters with
short dynamical timescales that have greater implications.  The force is with
this dark side.

\section{Acknowledgments}

The work involving the Numerical Action Method was done with Jim Peebles and 
Ed Shaya, while the work on the faint end of the luminosity function was with
Neil Trentham.  Support has been provided by JPL Contract 1243647, the STScI
award HST-GO-09162.01A, and the NSF award AST-0307706.  This paper is the 
extract
of a talk given on a special occasion for me, a workshop in my honor, and I
am indebted to all those that participated and made the event possible, 
especially Joss Bland-Hawthorn.


 \label{lastpage}

\end{document}